\documentclass[reprint,
bibnotes,
 amsmath,amssymb,
 aps,
floatfix,
]{revtex4-2}

\usepackage{graphicx}% Include figure files
\usepackage{dcolumn}% Align table columns on decimal point
\newcolumntype{d}[1]{D{.}{.}{#1}}
\usepackage{bm}% bold math
\usepackage{ulem}
\usepackage{xcolor}

\newcommand{\ddarrows}{\mathbin\downarrow\hspace{-.35em}\downarrow}
\newcommand{\downuparrows}{\mathbin\downarrow\hspace{-.35em}\uparrow}
\newcommand{\updownarrows}{\mathbin\uparrow\hspace{-.35em}\downarrow}

\begin{document}

\preprint{APS/123-QED}

\title{Near-critical dark opalescence in out-of-equilibrium SF$_6$}

\author{Valentina Martelli$^{1,*}$, Amaury Anquetil$^2$, Lin Al Atik$^2$, J. Larrea Jiménez$^1$, Alaska Subedi,$^{3,*}$, Ricardo P. S. M. Lobo$^{2,*}$ and Kamran Behnia$^{2,*}$}
\affiliation{$^1$ Laboratory for Quantum Matter under Extreme Conditions\\ Institute of Physics, University of São Paulo, 05508-090, São Paulo, Brazil\\
$^2$ Laboratoire de Physique et d'\'Etude des Mat\'eriaux (ESPCI - CNRS - Sorbonne Universit\'e)\\PSL University, Paris, 75005, France\\
$^3$ CPHT, CNRS, \'Ecole polytechnique, Institut Polytechnique de Paris, 91120 Palaiseau, France
}


%\affiliation{

%\affiliation{%


\date{\today}% It is always \today, today,

\begin{abstract}

 The first-order phase transition between the liquid and gaseous phases ends at a critical point. Critical opalescence occurs at this singularity. Discovered in 1822, it is known to be driven by diverging fluctuations in the density.
 %at the critical point. 
 During the past two decades, boundaries between the gas-like and liquid-like regimes have been theoretically proposed and experimentally explored. Here, we show that fast cooling of near-critical sulfur hexafluoride (SF$_6$), in presence of Earth's gravity, favors dark opalescence, where visible photons are not merely scattered, but also absorbed. When the isochore fluid is quenched across the critical point, its optical transmittance drops by more than three orders of magnitude in the whole visible range, a feature which does not occur during slow cooling. We show that transmittance shows a dip at 2eV near the critical point, and the system can host excitons with binding energies ranging from 0.5 to 4 eV.  The spinodal decomposition of the liquid-gas mixture, by inducing a periodical modulation of the fluid density, can provide a scenario to explain the emergence of this platform for coupling between light and matter. The possible formation of excitons and polaritons points to the irruption of quantum effects in a quintessentially classical context.

%The investigation of the supercritical region in the universal phase diagram of a fluid has recently regained attention. New theories have claimed a crossover separation between a liquid-like and gas-like behavior of the molecules, dubbed the Frenkel line, opening new perspectives on our understanding of a fluid's dynamics above and near the critical temperature $T_c$ and calling for new experiments. In this work, we report on the study of the optical behavior of near-critical sulfur hexafluoride. 
%SF$_6$ is a mostly anthropogenic molecule  that is recently raising growing concerns for its ability to efficiently absorb infra-red radiation and contribute significantly to global warming. 
%We have investigated the transmission of visible light through a SF$_6$ fluid and we discovered that SF$_6$ displays on earth wide black patterns when driven across its critical point upon fast cooling. Unlike critical opalescence, which is an efficient diffusive process, blackness points to full absorption of visible light, which we explain as a homogeneous emulsion-like fluid where highly-packed crystalline clusters with formed electronic bands are able to absorb photons with energy in the order of eV. 

%\begin{description}
%\end{description}
\end{abstract}

\maketitle

%\tableofcontents

\section{\label{sec:level1}INTRODUCTION}
The thermodynamic boundary between the liquid \cite{Proctor2021,Trachenko2023} and the gaseous states of matter ends at a critical point. Beyond this point, the substance becomes a supercritical fluid \cite{callen1998thermodynamics}, which is dense like a liquid and compressible like a gas \cite{carles2010brief_review_supercriticalFluid}, and it is employed in numerous applications, such as extraction, purification, or separation of chemical species across different industries \cite{brunner2010applications, knez2014industrial}. 

Critical phenomena \cite{domb1996critical} has been  studied for two centuries. As early as 1822, Charles Cagniard de la Tour observed the formation of a \textit{`nuage tr\`es \'epais'} (a very thick cloud) \cite{delatour1822}, during the liquefaction of the supercritical fluid.  Four decades later, Thomas Andrews \cite{andrews1869}, clearly identified the phenomenon: \textit{``...the surface of demarcation between the liquid and gas became fainter, lost its curvature, and at last disappeared. The space was then occupied by a homogeneous fluid, which exhibited, when the pressure was suddenly diminished or the temperature slightly lowered, a peculiar appearance of moving or flickering striae throughout its entire mass.'' } 

Dubbed critical opalescence, this phenomenon was found to occur universally at the critical end point of a liquid-gas transition \cite{domb1996critical}. Einstein \cite{Einstein1909} and, independently, Smolouchowski \cite{Smoluch1908} identified its origin by noting that fluctuations in the density (and therefore fluctuations in the refractive index) drastically amplify at the critical point. A few years later, Ornstein and Zernike (OZ) \cite{Ornstein1914} presented a more sophisticated treatment by introducing the pair-correlation function. A more elaborated version of the latter, taking into account corrections induced by interactions, was proposed half a century later by Fisher \cite{Fisher1964b}. This spectacular optical phenomenon \cite{Puglielli1970, chu1972critical_opalescence,zubkov1988critical_opalescence} is known to be one manifestation of the divergence of the correlation length at the critical point. The latter leads to singularities in a host of physical properties, including heat capacity, compressibility, viscosity and thermal conductivity \cite{Zappoli2014}. 

\begin{figure*}[ht!]
\includegraphics[width=1\linewidth]{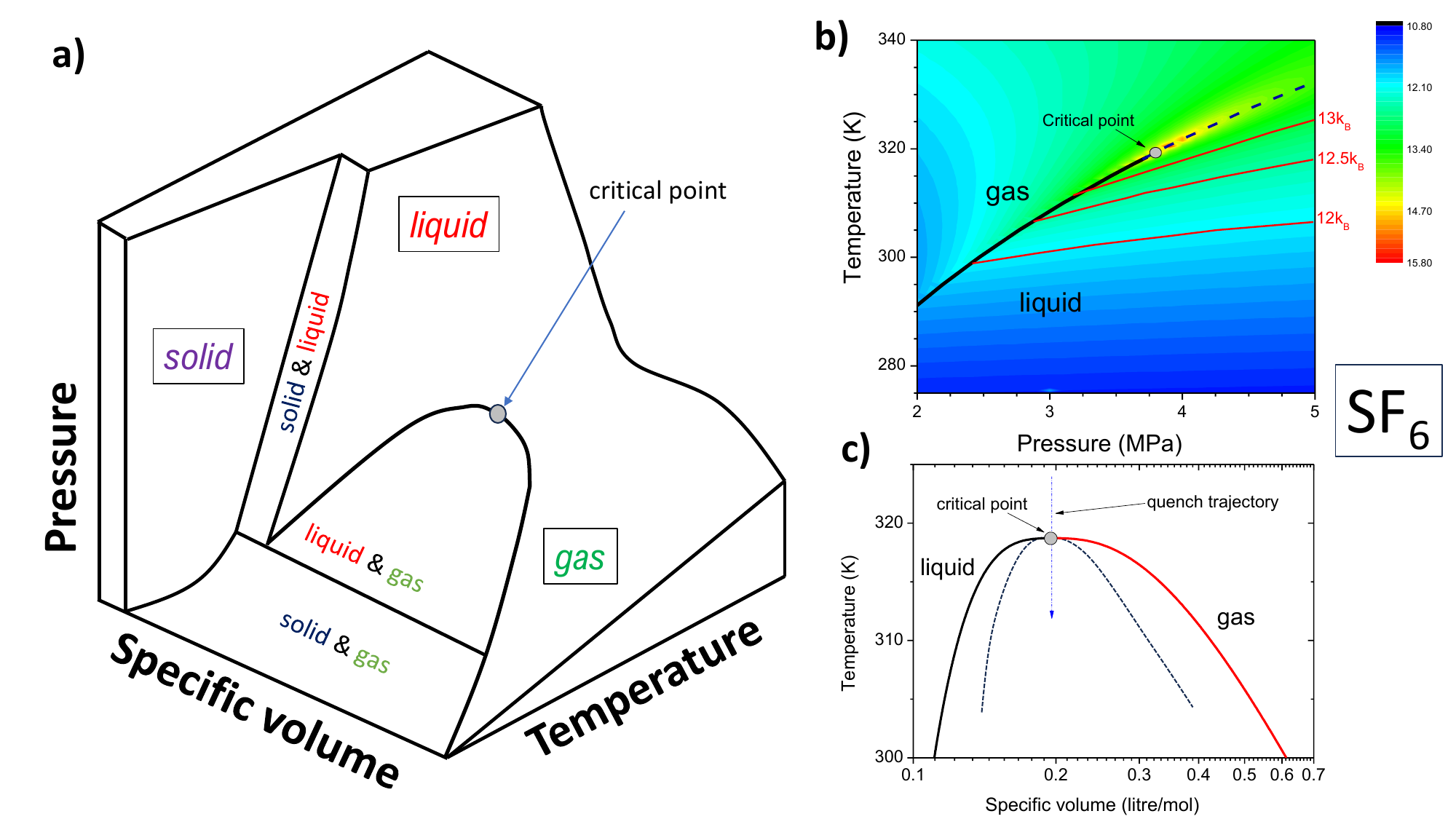} 
\caption{\textbf{Phase diagram of SF$_6$:} (a) The generic pressure, temperature, volume (p,V,T) Gibbs surface showing the three phases, their coexistence and their boundaries. (b) Color plot of the isochore specific heat of SF$_6$, $c_V$, in units of the Boltzmann constant near its critical point, according to the NIST database \cite{nist}. The melting curve (bold solid line) stops at the critical point and is followed by the Widom line (dashed line), identified by the peak in the specific heat data \cite{nist}). Three lines corresponding to  specific $c_V$ values are indicative of the [crossover] Frenkel line (See text). c) The liquid-gas boundaries of SF$_6$ in the (temperature, specific volume) plane \cite{nist}. The dashed line is an approximate sketch of the spinodal line separating unstable and metastable regions of liquid-gas coexistence. Quenching the fluid at constant volume will push it to the spinodal region. }
\label{fig:SF6-pd}
\end{figure*}

The idea that the supercritical fluid is not featureless has recently gained traction \cite{brazhkin2012separate, brazhkin2012_frenkel_line, brazhkin2013liquid_gas_frenkel_line, gorelli2013dynamics, trachenko2015collective_modes_liquid, Ruppeiner,Ploetz,Trachenko2021,COCKRELL20211} (See \cite{Simeoni2010,Bryk2017} for alternative views). Several boundaries inside the supercritical fluid, crossovers and not thermodynamic transitions (the Widom line, the Fisher-Widom line, and the Frenkel line) have been proposed.
%While they all refer to a passage in the behavior of the supercritical fluid from  gas-like to liquid-like, their focus is different. The Widom line \cite{Limei} is the curve along which the correlation length peaks, often tracked by a maximum in specific heat. The Fisher-Widom line \cite{Fisher2003} marks the boundary between two regimes of pair correlation function: it monotonically decays on the gas-like side and is oscillatory on the liquid-like side. In contrast to these two, the Frenkel line \cite{brazhkin2012_frenkel_line,brazhkin2013liquid_gas_frenkel_line} is based on the dynamic properties. It separates a rigid liquid, which allows solid-like oscillations in particle dynamics, and a non-rigid fluid, which does not allow such oscillations. Several experimental manifestations of the Frenkel line have been recently reported \cite{Trachenko2021,COCKRELL20211}. 

The dynamic response near the critical temperature is also affected by the competition between the critical slowing down of the heat transfer and the fast thermalization induced by the `piston effect'  \cite{carles2010brief_review_supercriticalFluid,Zappoli2014}, an adiabatic thermalization via acoustic waves, first invoked to explain unexpected features in micro-gravity experiments \cite{Onuki1990}.

With an easily accessible critical temperature ($T_c$ = 318.7 K), and critical pressure ($P_c= 3.76$ MPa), sulfur hexafluoride (SF$_6$) is a popular platform for the study of the critical point \cite{wilkinson1998equilibration_SF6, gorbunov2018dynamic_gravity, garrabos2007thermoconvectional} (Fig. \ref{fig:SF6-pd}).  During a study of near-critical SF$_6$ subject, we found that when it is quenched below its critical temperature (following the trajectory sketched in Fig. \ref{fig:SF6-pd}c), photons in the visible spectrum are absorbed and not merely scattered. The out-of-equilibrium fluid displays a black band of variable width, with details depending on the cooling rate and the relative orientation of pressure and temperature gradients. Measuring the transmittance across our 4~cm-thick fluid, we found that this blackness is concomitant with a large (four orders of magnitude) attenuation of the transmission coefficient and the appearance. For both  slow cooling and fast cooling, we  detect an absorption peak at 2~eV.

Critical opalescence, which refers to an amplified turbidity in a fluid kept at thermodynamic equilibrium and pushed to its critical point, is driven by the amplification of Rayleigh and Brillouin scattering of light. It has been experimentally quantified in the presence \cite{Puglielli1970} and in the absence \cite{Lecoutre2009} of gravity. Puglielli and Ford \cite{Puglielli1970} verified the OZ theory and quantified the correlation length of SF$_6$ from their data. In their micro-gravity experiment, Lecoutre \textit{et al.}  \cite{Lecoutre2009} quantified turbidity to within a fraction of millikelvin of the critical temperature and detected deviations from the OZ theory associated with particle-particle interaction, as proposed by Fisher \cite{Fisher1964b}. 

In contrast with these studies, we find that the out-of-equilibrium fluid during a quench becomes black and not merely turbid. One cannot explain this blackness, which is destroyed by reducing the cooling rate and approaching the thermodynamic equilibrium, within the framework of the OZ theory. We argue that the Coulomb attraction between electrons and holes is strong in our context providing a context for the formation of excitons \cite{Frenkel1931,Wannier1937} and their interaction with photons in the visible spectrum. This constitutes a first step towards understanding the experimental observation, which still lacks a  solid theoretical account.

A promising framework for our study is the Cahn-Hillard theory of spinodal decomposition~\cite{CAHN1961795,Cahn1965,Elliott1989}, which treats  binary solids or liquid solutions quenched out of equilibrium. In our experiment, we force our fluid to a point in the phase diagram where the two (gaseous and liquid) phases have different specific volumes at thermodynamic equilibrium. The Cahn-Hillard theory postulates that within the spinodal region (see the dashed lines in Fig. \ref{fig:SF6-pd}c), because of the absence of a thermodynamic barrier, the decomposition is governed by  an `uphill' diffusion. This leads to the emergence of spinodal nanostructures in metallic alloys \cite{FINDIK2012131} and in polymer blends \cite{CABRAL20181}. In our case, the spatial modulation of the density generated by the quench may favor a type of  light-matter coupling reminiscent of polaritions \cite{Basov2021} in  solid state heterostructures.  

%Our measurements confirm that the supercritical fluid is not featureless \cite{COCKRELL20211} and that the dynamic response to temperature and pressure gradients generates complex patterns \cite{Cross1993} driven by non-trivial dynamics of light-matter interaction near the critical point.

\section{\label{sec:EXPERIMENTAL} EXPERIMENTAL}

 Fig.~\ref{fig:SF6-pd}(b) presents a color map of the isochore specific heat (c$_V$) of SF$_6$ near its critical point in units of k$_B$, according to the NIST database \cite{nist}. The boiling line separating the liquid and the vapor phases ends at the critical point, where the Widom line starts. Here, it is plotted by tracking the experimentally resolved \cite{nist} peak of the specific heat.  The  Frenkel line, separating the rigid and the non-rigid fluids  \cite{brazhkin2012_frenkel_line,Proctor2019} is expected to be at lower temperature starting in the liquid state and eventually becoming parallel to the Widom line at higher temperature and pressure. 

\begin{figure}[ht!]
\includegraphics[width=1\linewidth]{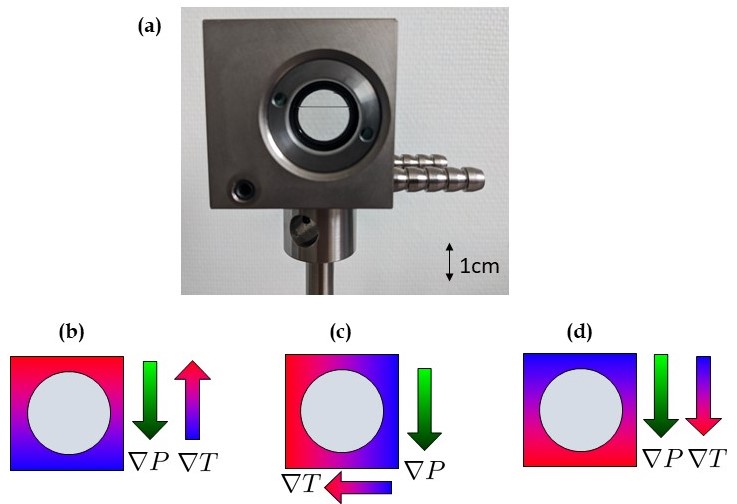} 
\caption{\textbf{The set-up and the three configurations:} (a)  A picture of the chamber containing SF$_6$. (b), (c) and (d) represent the three configurations explored in this work with different orientations of the temperature gradient with respect to the pressure gradient.}
\label{fig:setup}
\end{figure}

Our isochore investigation, along the trajectory sketched in  Fig.~\ref{fig:SF6-pd}(c), used a pressure chamber filled with liquefied SF$_6$ gas commercially obtained from Leybold Didactic GMBH (Fig. \ref{fig:setup}a) \cite{Leybold}. 

\begin{figure}[ht!]
\includegraphics[width=1\linewidth]{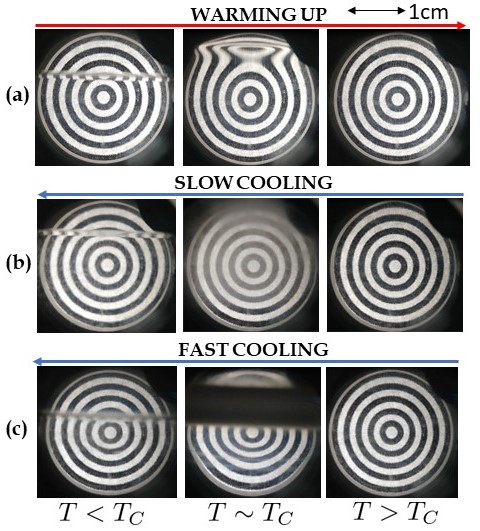} 
\caption{\textbf{Warming, slow cooling and fast cooling:} Concentric circles were put behind the back window of the chamber to show the evolution of transparency, with chamber in the $\nabla T$ $\downuparrows$ $\nabla P$ configuration. (a) Evolution across the critical temperature during a Warming process from $T<T_c$ to $T>T_c$. The phase separation fades away gradually, and the fluid remains transparent. (b) A slow cooling process in the reverse direction (from $T>T_c$ to $T<T_c$) reveals opacity near $T\sim T_c$. Despite turbidity, circles remain visible. (c)  A dark band shows up during a fast cooling process (a quench across $T_c$).}
\label{fig:fig3}
\end{figure}

A platinum sensor placed into a lateral cavity allowed monitoring of the approximate temperature $T_{\it Pt}$ of the chamber. The latter is provided with two optical windows and inner pipes to circulate cooling water. The setup includes a light bulb in front of one of the two optical windows and a standard optical camera in front of the other one. The fluid images were recorded during the experiment. The circulation pipes were connected to a water bath with a controlled temperature. The arrangement allowed for setting a reproducible, stable, and uniform temperature of the chamber above $T_c$ by heating. 

The water circulation cools one side of the chamber before the other, generating a thermal gradient. Using an infrared camera, we verified (see the supplemental material \cite{SM}) the orientation of the temperature gradient as depicted in Fig. \ref{fig:setup} for the three configurations. Since the cavity of the sensor is placed on the cold side of the chamber, in the following, the sensor temperature, $T_{\it Pt}$, refers to the colder end of the temperature gradient. The chamber did not allow for modifications to host sensors inside itself. Therefore, the amplitude of the temperature gradient inside the fluid is known with limited accuracy.

The chamber was fixed on a small homemade optical table with dumping supports decoupling the fluid from external vibrations and could be oriented along three orientations. The pressure gradient  ($\nabla P$) due to gravity points downward and,  using the critical mass density of SF$_6$ ($\rho_c = 742$ kg.m$^{-3}$) is estimated to be $\frac{dp}{dz}\simeq 7.3$ Pa.mm$^{-1}$. 

We studied three different configurations where the temperature gradient across the chamber ($\nabla T$) and the pressure gradient ($\nabla$ P)  were either perpendicular ($\nabla P \perp \nabla T$), anti-parallel ($\nabla P$ $\updownarrows$ $\nabla T$) or parallel ($\nabla P$ $\ddarrows$  $\nabla T$), as depicted in panels (b) to (d) of Fig.~\ref{fig:setup}. 

A measurement protocol was defined and reproduced for each configuration: the chamber was heated up to 325 K and then cooled down at different rates. The fluid is in the supercritical state and is transparent [see right panels of Fig.~\ref{fig:fig3}(a), (b) and (c)]. Upon cooling down phase separation occurs and the meniscus was observed by the camera. Frames were recorded with a fixed sampling frequency together with the corresponding $T_{\it Pt}$. The recorded images were processed through a Python routine comparing each frame by its precedent in order to quantify the observed variation as a function of time and $T_{Pt}$. The frames were first translated in a scale of greys and then compared using the mean squared error (MSE) index, defined as  
\begin{equation}
\label{MSE}
 {\it MSE}=\frac{1}{mn}\sum_{j=0}^{n-1} \sum_{i=0}^{m-1} \left[ I(i,j)-I_0(i,j)\right]^2
\end{equation}
where $I(i,j)$ represents the intensity of the $(i,j)$ pixel in a scale of greys quadratically summed over all the matrix columns and rows of the image with respect to the $I_0(i,j)$ calculated at the initial conditions; $(m,n)$ are the rows and columns that span the full matrix of pixels. 

Optical spectroscopy transmission measurements complemented the image recording experiments. We compared the transmittance across the fluid when it was cooled very slowly to when it was quenched rapidly below the critical point. Transmittance spectra in the visible range were collected from 1100 to 350~nm (1 to 3.5~eV) in an AvaSpec 2048-14 optical fiber dispersive spectrometer with a resolution of 2~nm. We utilized a deuterium-arc combined with a halogen lamp to cover this range. The spectrometer diffraction grating, combined with the sensitivity of the CCD detector, limits the low-energy range of the spectrum. The high-energy cut-off comes from the two glass windows of the chamber. The transmission of the liquid and the supercritical phases was normalized by the transmission of the gaseous phase measured at 315~K. All measurements were collected starting from the supercritical phase at 320~K. In the slow-cooling experiment, measurements were performed with 1000 averages of 2.17~ms-integration-time spectra. During the quench (i.e. fast cooling) experiment,  the spectra-acquisition rate was decreased to 100 averages. All the spectra were collected in the ($\nabla P \perp \nabla T$) configuration.

\section{\label{sec:RESULTS} RESULTS}
Distinct behaviours were observed during warming up and cooling down [Fig. \ref{fig:fig3} (a), (b), and (c)]. Below $T_c$, the gas and liquid phases were visibly distinct. With warming, the meniscus slowly moves upward until it vanishes and leaves the full space for the supercritical phase. 

Figure \ref{fig:fig3} shows a comparison of the evolution of transparency of SF$_6$ during a fast and a slow cooling process. In contrast with the quasi-static process, shown in Fig.~\ref{fig:fig3}(b) (See also Fig.1 in ref. \cite{Zappoli2014}), in the fast-cooling process a black horizontal pattern abruptly appears  [Fig. \ref{fig:fig3}(c)] and then gradually fades away and is replaced by a meniscus separating the two distinct phases. The patterns depend on the cooling rate and configurations.

\subsection{Visual recording of emergent darkness in the three configurations}
\begin{figure*}[ht!]
\includegraphics[width=0.8\linewidth]{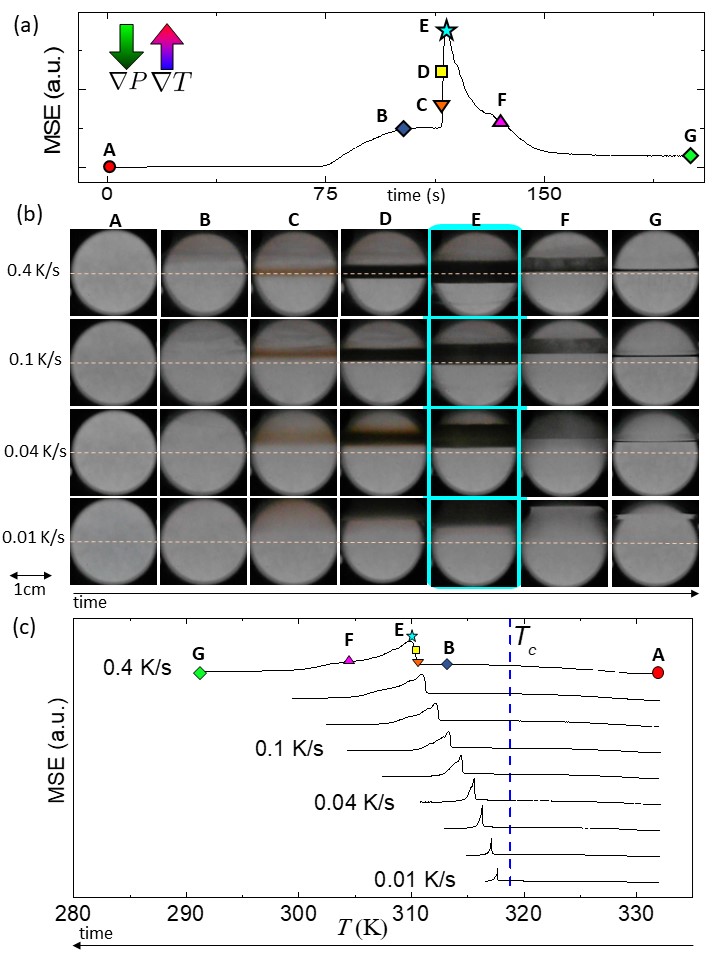} 
\caption{\label{fig:antiparallel} \textbf{$\nabla T$ $\downuparrows$ $\nabla P$ configuration:} (a) Time evolution of the MSE index as a function of time.  Frames at selected points are labelled from \textbf{A} to \textbf{G}. (b) Comparison of the \textbf{A} to \textbf{G} frames for different cooling rates (0.4K/s to 0.01K/s). The blue background (behind the \textbf{E} frames) points to the maximum in MSE for each data set. (c) Temperature dependence of the similarity index (MSE). The red arrow points to the peak corresponding to the maximum extension of the black pattern. Data are shifted downward to allow visualization. The blue dotted line shows the position of the critical temperature $T_c$.}
\end{figure*}

\begin{figure*}
\includegraphics[width=0.8\linewidth]{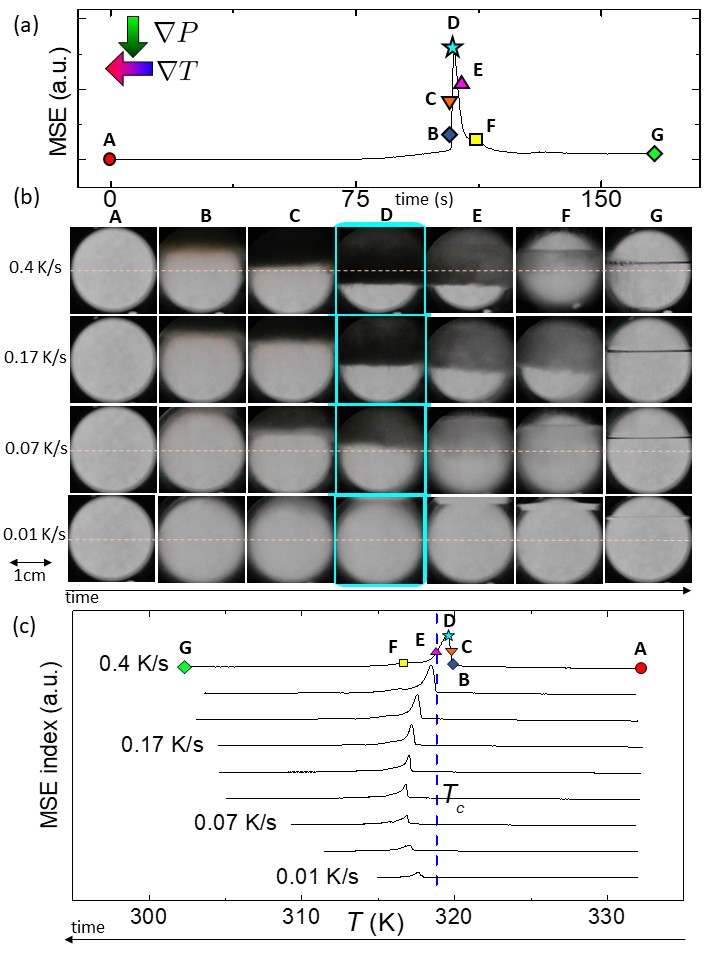} 
\caption{\label{fig:perpendicular} \textbf{$\nabla T$ $\perp$ $\nabla P$ configuration:} (a) Time evolution of the MSE index as a function of time.  Frames at selected points labelled from \textbf{A} to \textbf{G}. (b) Comparison of the \textbf{A} to \textbf{G} frames for different cooling rates (0.4K/s to 0.01K/s). The blue background (behind the \textbf{D} frames) points to the maximum in the MSE index for each data set. (c) Temperature dependence of the similarity index (MSE). The red arrow points to the peak corresponding to the maximum extension of the black pattern. Data are shifted downward to allow visualization. The blue dotted line shows the position of the critical temperature $T_c$.  }
\end{figure*}

\begin{figure*}
\includegraphics[width=0.8\linewidth]{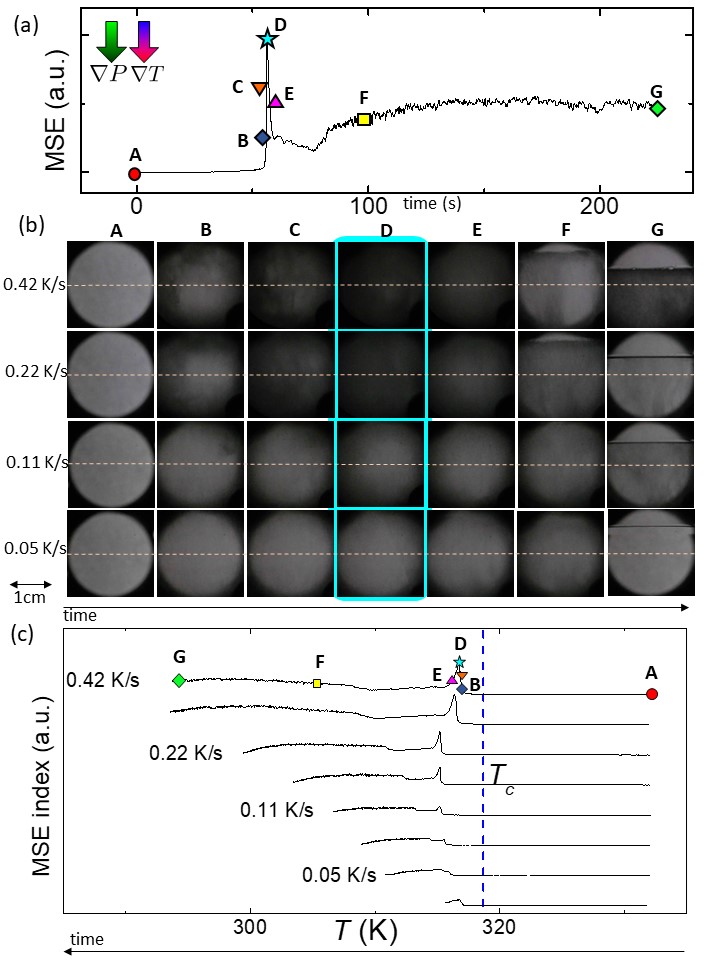} 
\caption{\label{fig:parallel} \textbf{$\nabla T$ $\ddarrows$ $\nabla P$ configuration:} (a) Time evolution of the MSE index as a function of time. Frames at selected points labelled from \textbf{A} to \textbf{G}. (b) Comparison of the \textbf{A} to \textbf{G} frames for different cooling rates (0.4K/s to 0.04K/s). The blue background (behind the \textbf{D} frames) points to the maximum in MSE for each dataset. (c) Temperature dependence of the similarity index (MSE). The red arrow points to the peak corresponding to the maximum extension of the black pattern. Data are shifted downward to allow visualization. The blue dotted line shows the position of the critical temperature $T_c$.}
\end{figure*}

The results obtained for the three configurations are presented in the next three figures (\ref{fig:antiparallel}, \ref{fig:perpendicular}, and \ref{fig:parallel}). We proceed to discuss each configuration.

\textbf{The anti-parallel configuration.} Figure \ref{fig:antiparallel} shows the data for this configuration ($\nabla P$ $\downuparrows$ $\nabla T$). Panel (a) shows the temporal evolution of the MSE-index (Eq. \ref{MSE}) extracted from processing the image data together with the corresponding frames at selected instants of times (labelled from \textbf{A} to \textbf{G}) during a cooling rate of 0.4 K/s. The MSE index sharply peaks when the black pattern occupies the widest region. It is, therefore, a reliable quantifier of the black pattern evolution. 

Figure \ref{fig:antiparallel}(b) shows fluid picture frames during the cooling process. One can see that the temporal evolution of the blackness depends on the cooling rate. The evolution of the MSE index as a function of $T_{Pt}$ is shown in Figure \ref{fig:antiparallel}(c). For fast cooling, the sharp peak occurs well below  the critical temperature. As the cooling rate slows down, the temperature at which the MSE index peaks and the blackness appears rises and becomes closer and closer to the critical temperature.

Frames \textbf{A} to \textbf{G} in Figure \ref{fig:antiparallel}(b) reveal a sequence of patterns. In frame \textbf{B}, there is a weak shadow on the upper side of the chamber; in \textbf{C}, a brownish horizontal pattern shows up and then becomes a dark black band in \textbf{E}. For fast cooling, the black horizontal band is centred in the chamber. As the cooling rate decreases, it shifts up toward the upper side of the chamber. At the lowest cooling rates ($<0.2~K/s$), the black region shows up only from above.

\textbf{The perpendicular configuration.} Fig.~\ref{fig:perpendicular} shows the results for the $\nabla P$ $\perp$ $\nabla T$ configuration. The temporal evolution of the MSE index is shown in Fig.~\ref{fig:perpendicular}(a). Its thermal evolution is displayed in Fig.~\ref{fig:perpendicular}(c). They both differ from the previous configuration. In particular, the temperature at which the MSE index peaks is barely lower than the critical temperature and shows a much weaker dependence on the cooling rate. In picture frames of the fluid during the cooling [Fig.~\ref{fig:perpendicular}(b)], one can see qualitative differences with the previous configuration. A black region appears from above (frames \textbf{B} and \textbf{C}). For the fastest cooling rate, this black region gradually covers more than half of the surface. It then becomes gradually transparent and then the phase separation line emerges. In this configuration, when the cooling rate is lower than 0.02~K/s, the dark pattern is almost absent (see \textbf{F}). It is replaced by a blurred frame with no blackness. 

\textbf{The parallel configuration.} The data for the last explored configuration ($\nabla T$ $\ddarrows$ $\nabla P$) is shown in Fig.~\ref{fig:parallel}. The time dependence of the MSE index [Fig.~\ref{fig:parallel}(a)], as well as its temperature dependence [Fig.~\ref{fig:parallel}(c)], show subtle differences compared to previous configurations. In particular, well after the peak and the separation between phases [between points \textbf{F} to \textbf{G} in Fig.~\ref{fig:parallel}(a)], there is a large noise indicating persistent turbulence. According to the cooling frames [Fig.~\ref{fig:parallel}(b)], the blackness is not spatially confined and occupies the whole frame. At the slowest cooling rates, there is no clear blackness at all and the MSE, instead of showing a sharp peak, displays a broad maximum near $T_c$ [Fig.~\ref{fig:parallel}(c)]. It is followed by a turbulent process below the critical temperature. 

Despite the identical cooling protocol, salient features of the collected data differ in the three cases; a detailed comparison of the evolution of the maximum of the MSE-index and the $T_{Pt}$ as a function of cooling rate is reported in the Supplemental material \cite{SM}. 

\subsection{Transmittance spectra}
\begin{figure*}[ht!]
\includegraphics[width=0.85\linewidth]{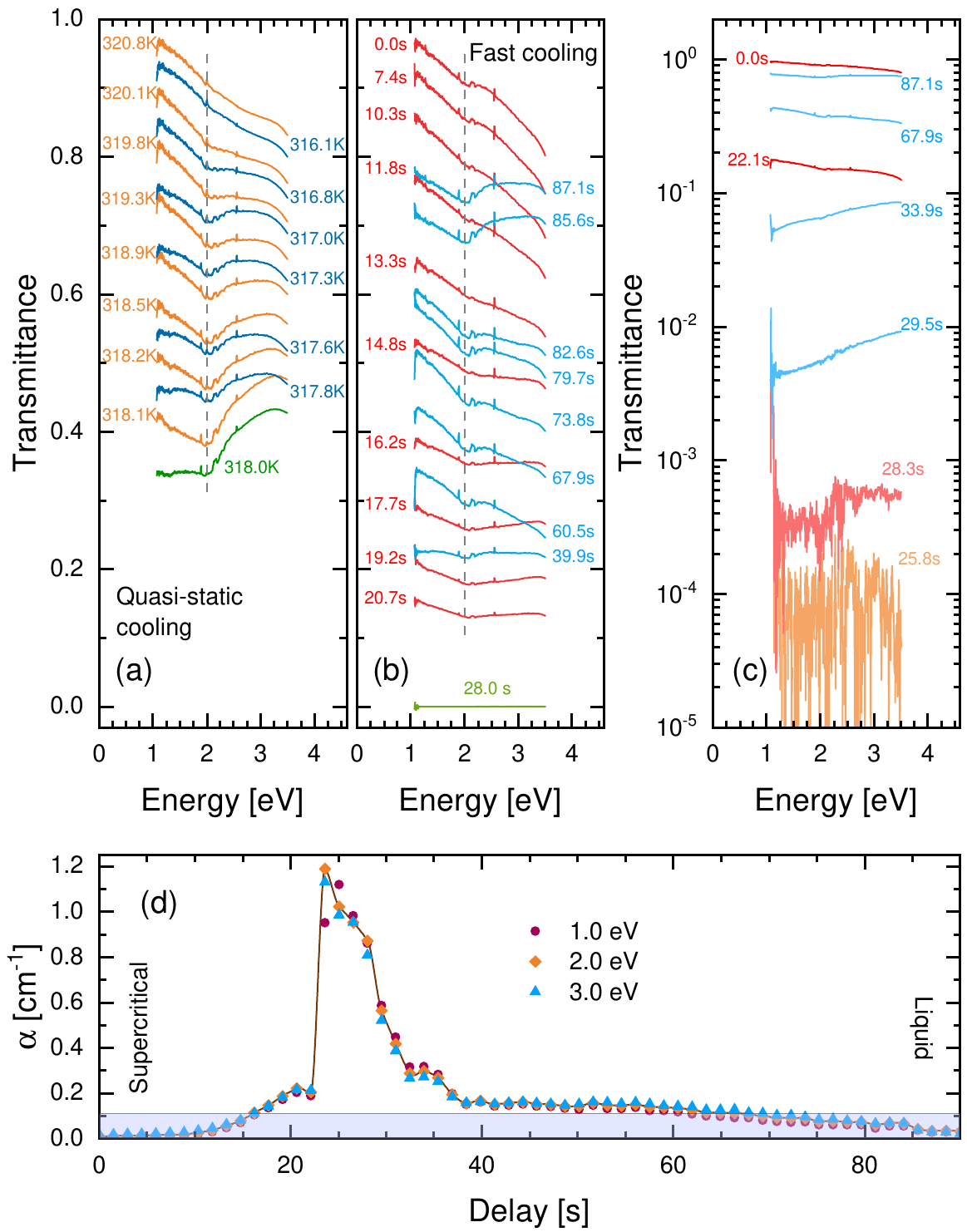} 
\caption{\label{fig:transmission} \textbf{Transmittance of the liquid or supercritical phases with respect to the gas phase: }(a) Slow cooling from the supercritical (orange curves) to the liquid (blue curves) phase. The green curve shows the minimum value for the transmission. The top two curves, at 320.8~K and 316.1~K, show the transmission of the supercritical and the liquid phases, respectively. A small dip, absent from the end spectra, appears around 2~eV (shown by the vertical dashed line). (b) Fast (0.3~K/sec) cooling of the sample starting from 320~K over 90~s. The same 2~eV-dip from the quasi-static measurements appears in the mixed phase. However, contrary to the quasi-static measurement, the sample becomes fully opaque with a transmittance close to zero 28~s after cooling starts. (c) Fast-cooling transmittance in logarithmic scale. The lowest value for the transmittance is at least 4 orders of magnitude smaller (value limited by the detector sensitivity) than the transmittance of either the supercritical or the liquid phase. (d) Time-dependence of the absorption coefficient at selected photon-energies. The solid line is a guide to the eye. The delay time is measured from the supercritical phase at $t = 0$. The shaded area illustrates the range of values for $\alpha$ in the quasi-static measurements of panel (a).}
\end{figure*}

In order to quantify the full opacity effect, we measured the transmittance of the sample in the visible range in the perpendicular configuration for two extreme cooling rates (i.e. $\sim$ 0.02 K/s and $\sim$ 0.3K/s).  The data is shown in Fig.~\ref{fig:transmission}. 

Panel (a) shows a selection (from spectra at 29 different temperatures) of measurements during a slow-cooling process, starting from the supercritical phase at 320~K. This process corresponds to images shown in the bottom row of Fig.~\ref{fig:perpendicular}b. The overall transmittance value decreases with temperature (orange curves), down to 318~K (green curve) then increases (blue curves) back to a value close to the initial measurement. The transmittance of the supercritical (320.8~K) and liquid (316.1~K) are very similar. There are two important features in this data. First, the overall transmittance decreases to about 35\% of its maximum value. Even if, visually, this is barely noticeable and easily hidden by the turbid aspect of the mixed-phase, the transmittance does decrease by a large amount. A second, more interesting feature, is the appearance of a marked dip around 2~eV in the transmittance. The amplitude of this dip is small, suggesting either a very weak absorption from the whole sample or, alternatively, a sparse distribution of highly absorbing regions. Remarkably, this marked 2~eV absorption feature appears only in the mixed state. It is absent from both the supercritical and the liquid phases. 

Panel (b) shows the transmittance when the sample is cooled at a fast rate of about 0.3~K/sec. Here the temperature is much less well defined as there is a large thermal gradient inside the chamber. We took 62 spectra over 90~s. The data is labelled with the delay after the cooling starting time. The initial overall behavior is the same. Starting from the supercritical phase, the transmittance decreases (red curves) and a dip in the vicinity of 2~eV appears. However, instead of having the transmittance saturating at about 35\% of the initial value, here the sample goes fully opaque as shown by the green curved taken 28~s after cooling begins. The transmittance then increases again (blue curves) and the 2~eV dip remains well marked. The large thermal gradient in the chamber implies that we are far from an equilibrium state. As a consequence, this panel does not show the transmittance fully recovering to the liquid state transmittance. Nevertheless, if we wait a few minutes after cooling stops, the transmittance spectrum comes back to the regular spectrum of the liquid state. 

In panel (c), we plot the fast cooling transmittance on a logarithmic scale. This figure shows that the system reaches full opacity faster than it recovers its transparency. In addition we see that, in the fast-cooling measurements, the transmittance is at least 4 orders of magnitude smaller than the one from the supercritical or the liquid phases. Note that the $10^{-5} - 10^{-4}$ range shown in this figure is a limitation of the CCD detector sensitivity rather than a measurement of the sample transmittance.

The absorption coefficient can be defined as $\alpha = - d^{-1} \log_{10} T$, where $d$ is the sample thickness  and $T$ the transmittance (shown in panels (a)--(c)). Panel (d) shows the time dependence of $\alpha$, taking $d = 4$~cm, at 3 selected energies -- below (1~eV), above (3~eV), and at the minimum of the absorption peak (2~eV). The time dependence is essentially the same for the 3 energies. The fluid is in the supercritical phase at the delay time $t = 0$. We notice that the absorption coefficient increases slowly to about the 22~s mark where a steep rise gives a five-fold increase to $\alpha$. It then still shows a sharp but slower decrease to the 40~s mark. From that point on, it takes a long time (a few minutes) for the transmittance to reach its liquid state equilibrium phase. This behavior is qualitatively the same as the one shown in Figs.~\ref{fig:antiparallel}, \ref{fig:perpendicular}, and \ref{fig:parallel} (note that the time scale is reversed in the data shown as a function of temperature). The shaded area in this panel shows the range of values that $\alpha$ takes in our quasi-static measurement, which is, at least, one order of magnitude smaller than the maximum $\alpha$ in fast-cooling spectra.

\subsection{Previous observations of possibly related phenomena}
We have found several reports in scientific literature reporting on darkness observed near a critical point.  Garrabos \textit{at al.} studied supercritical CO$_2$ under micro-gravity ($10^{-4}$ times less than terrestrial gravity) \cite{garrabos1992observation_spinoidal_decomposition_CO2} and observed a pattern of interconnected domains and rapid density fluctuations whose origin was not identified. In a related study, Guenoun and co-workers \cite{garrabos1992observation_spinoidal_decomposition_CO2, guenoun1993thermal_cycle_CO2} found black stripes in the images recorded slightly above $T_c$ and attributed it to density gradients induced by a gravity of 2g. Dark granular domains (each containing turbulent activity) were also observed. Ikier \textit{et al.}~\cite{ikier1996optical_SF6} monitored cooling across $T_c$ in SF$_6$, and observed droplets domains with dark circular regions constituting the meniscus between the liquid-gas separation at each droplet. The formation of such droplets under reduced gravity was studied in detail \cite{beysens1997kinetics_droplet_coalescence,oprisan2014dimple}.

\section{\label{sec:DISCUSSION} DISCUSSION}
Critical opalescence refers to the gradual enhancement of turbidity near the critical point when the latter is approached in thermodynamic equilibrium. Experiments both on Earth \cite{Puglielli1970} and in space \cite{Lecoutre2009} have documented this phenomenon. Our observation is qualitatively distinct. The blackness observed occurs when the two phases are separated and are amplified with an increasing cooling rate. 

The Ornstein-Zernike theory of critical opalescence invokes Rayleigh and Brillouin scattering and the Lorenz-Lorentz relation \cite{Kragh_2018} linking the refraction index to density. Puglielli and Ford have shown that it leads to the following expression for the scattering rate \cite{Puglielli1970}:
\begin{equation}
\tau_{0}= \frac{8}{3} \frac{\pi^3}{\lambda^4} \rho^2 \left(\frac{\partial \epsilon_r}{\partial \rho}\right)_T^2k_BT_c \kappa_T \quad .
\label{Eq:turbidity}
\end{equation}
Here, $\lambda$ is the photon wavelength,  $\epsilon_r$ is the electric permittivity of the medium, $\rho$ is its density and $\kappa_T$ is the isothermal compressibility. $\tau_{0}$  has the dimensions of the inverse of length. Its temperature dependence is set by the temperature dependence of the compressibility which diverges at the critical point. The experimental validity of equation \ref{Eq:turbidity} near the critical point of SF$_6$ was experimentally confirmed on Earth, for $(T-T_c)>0.03$ K \cite{Puglielli1970}, and by very precise micro-gravity experiments to within $(T-T_c) \approx 0.3$ mK \cite{Lecoutre2009}. 

Our out-of-equilibrium data cannot be accounted for by this equation, which associates the amplification of $\tau_0$ with the enhancement of the isothermal compressibility, ($\kappa_T$), and its divergence with the approach of the critical point.  The wavelength of visible photons ($400~\textrm{nm} < \lambda < 700~\textrm{nm}$) and the critical temperature puts $\frac{8\pi^3}{3}\frac{k_BT_c}{\lambda^4}$ in the range of $10^7-10^8$ J.m$^{-4}$. The order of magnitude of $\rho\frac{\partial \epsilon_r}{\partial \rho} \simeq \frac{\partial \epsilon_r}{\partial ln\rho}$ is bounded by the Lorentz-Lorenz relation. In this context,  Eq. \ref{Eq:turbidity} would attribute the magnitude of $\tau_0$ $\approx 1~\textrm{cm}^{-1}$ (seen during a fast cooling) to an unrealistically large compressibility (1 MPa $^{-1}$). It is unlikely that the bulk modulus of the fluid becomes suddenly as small as 1 MPa because of fast cooling. The bulk modulus supercritical SF$_6$ (at one degree off the critical temperature) is $\approx$ 60 MPa \cite{Marekevitch}.  

The spectral response brings an important piece to the puzzle. The liquid and supercritical phases are both essentially colorless transparent suggesting featureless transmittance in the visible range. Indeed, the transmittance at 320~K and 316~K of Fig.~\ref{fig:transmission}(a) do not show any absorption lines. The transparency of the end phases led to previous optical studies that looked at integrated or monochromatic transmittance measurements \cite{Puglielli1970,Lecoutre2009}. The absence of absorption lines in the spectra impelled, naturally, a description of the critical opalescence in terms of light diffusion only. 

However, in our measurements, near the critical point and only there, an optical absorption band, hence electrical-dipole-coupled transition, appears around 2~eV. %This is a feature that only exists when the liquid droplets are immersed in a gas matrix (or vice-versa). Even though diffusion (geometrical) processes are important in the disordered, non-equilibrium, mixed state. 
The presence of an absorption peak shows that quantum mechanical processes (optical transitions)  are at work. 

In reasonably transparent media, the absorption coefficient is dominated by $\varepsilon^{\prime\prime}$, the imaginary part of the dielectric function. 
A quantum-mechanical dipole-transition formalism leads to~\cite{Tanner2019}:
\begin{equation}
\varepsilon^{\prime\prime} (\omega) = \frac{4 \pi n e^2}{m} \frac{\gamma \omega f_0}{(\omega_0^2 - \omega^2)^2 - \gamma^2 \omega^2} \quad ,
\end{equation}
where $\omega_0$ is the dipole-excitation resonance frequency, $\gamma$ the resonance line-width (inverse lifetime). The parameter $f_0 = 
\frac{2 m \omega_0}{\hbar}|\langle 0|x|n \rangle|^2$, dubbed the oscillator strength, is proportional to the induced dipole matrix element between the ground and excited states. It is fair to wonder what generates $\omega_0\simeq 2 eV$ in our case.
%}

The vibrational or rotational energy scales of single molecules are too low. The largest Raman-active mode of a SF$_6$ molecule is only $\approx 0.096$ eV \cite{RUBIN1978254,aboumajd1979analysis_IR_SF6} and the strongest infrared absorption (which makes SF$_6$ a powerful green house gas) occurs near 0.117 eV \cite{boudon2014_SF_IR}.  

On the other hand, the electronic band gap of the solid state is as large as 7 eV (See the details in the supplemental material). This roughly quantifies the hopping energy of electrons across adjacent (van der Waals bound) SF$_6$ molecules and is expected to be relevant in a liquid phase of similar density. After all, optical gaps in crystalline and amorphous silicon have comparable amplitudes \cite{Knief1999}. We also note that the experimentally resolved band gap in solid and liquid Xe are close to each other ($\approx$ 9eV) \cite{Steinberger1973}. More generally, the phonon theory of liquid thermodynamics \cite{Bolmatov2012} has recalled the existence of solid-like behaviour in liquids in many aspects.

\begin{figure}[ht]
\includegraphics[width=1\linewidth]{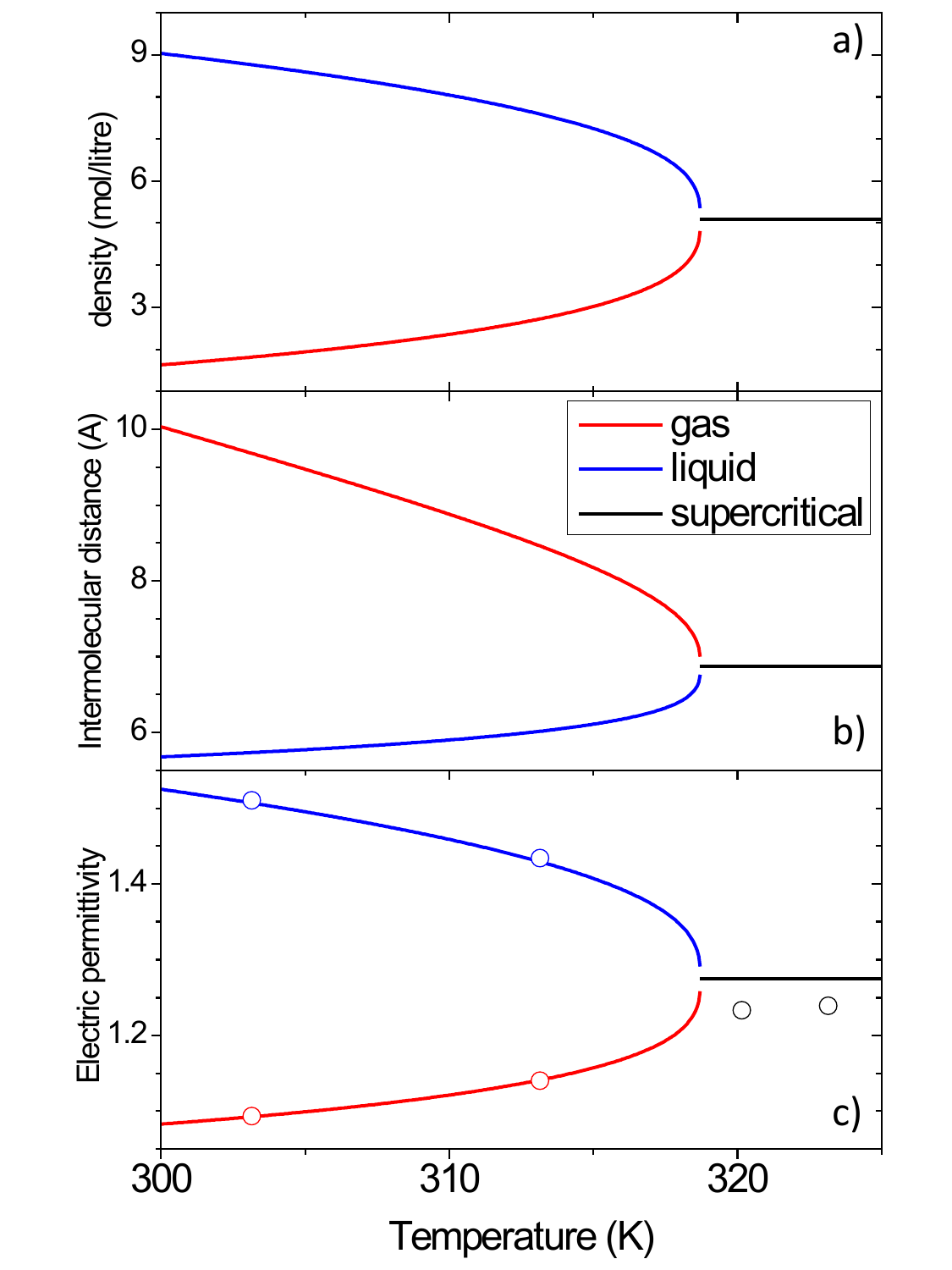} 
\caption{\textbf{Density, intermolecular distance and electric permittivity:} a) The isochore density of SF$_6$ as a function of temperature near the critical point ref. \cite{nist}. Below the critical temperature, the liquid and the gaseous phases have different densities. b) The intermolecular distance calculated from the density for the three phases. c) The electric permittivity, $\epsilon_r$, of SF$_6$ as a function of temperature near the critical point for the three phases calculated from the density, using the Clausius-Mossotti expression and the polarization of the SF$_6$ molecule.  Empty circles show the experimentally measured electrical permittivity of SF$_6$ \cite{Kita1994}. }
\label{Fig:epsilon} 
\end{figure}

\begin{figure*}[ht]
\includegraphics[width=0.95\linewidth]{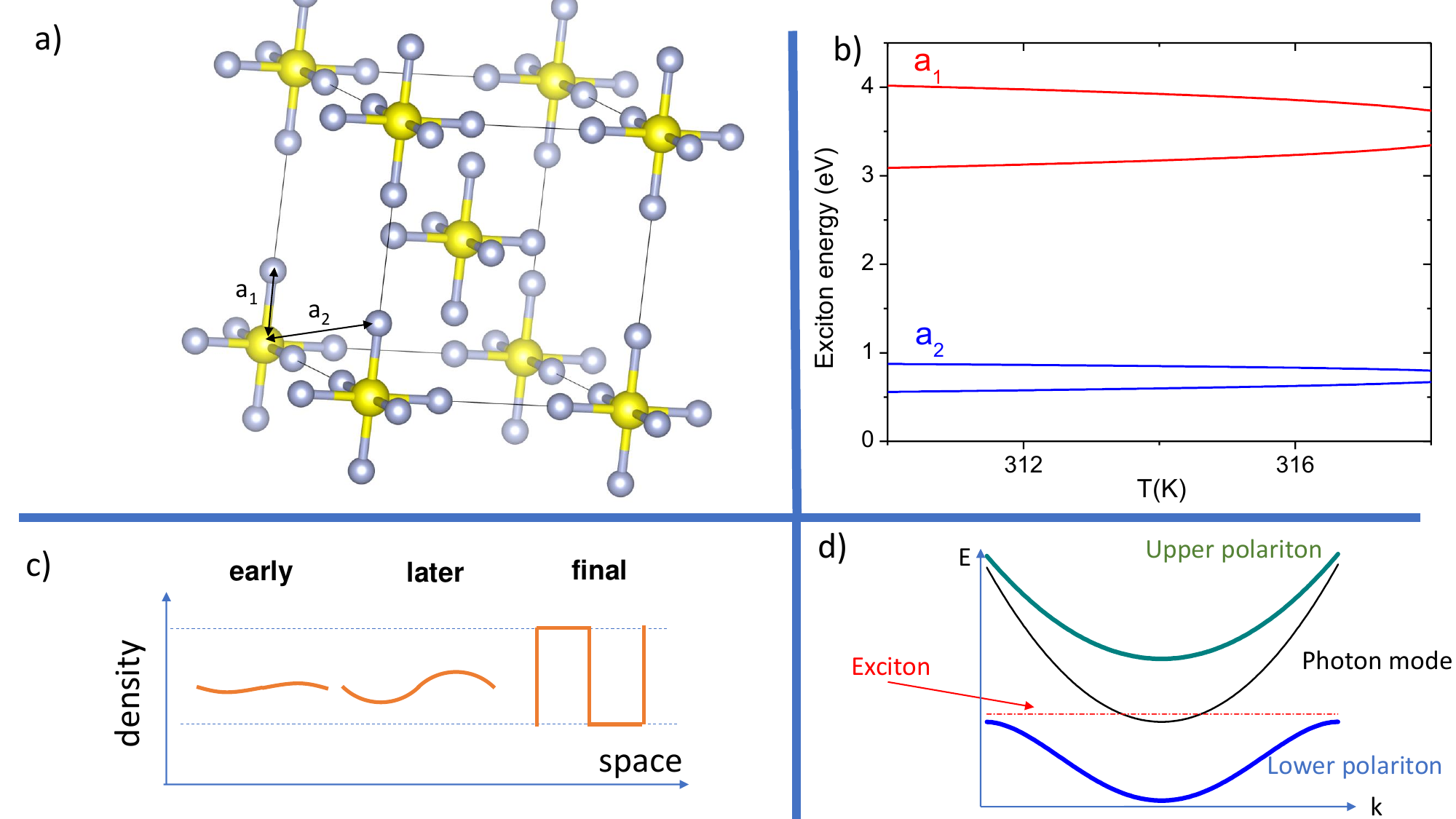} 
\caption{ \textbf{Excitons and their binding energies in SF$_6$:}  a) SF$_6$ molecules in a body cubic centered configuration. Golden (silver) spheres represent S (F).  $a_1$ and $a_2$ designate two radii for electron-hole pairs. The first length, $a_1$, is fixed, but $a_2$ increases with increasing intermolecular distance. b)  Upper and lower boundaries for exciton energies according to Eq. \ref{Eq:Exc1}  for $d_{eh}=a_1$ and for $d_{eh}=a_2$ and given the variation of intermolecular distance with temperature. c) A sketch of the evolution of density inhomegeneity during spinodal decomposition \cite{FINDIK2012131}. The density gradient gently rises and eventually saturates to its full contrast. d)  The dispersion of an exciton, a cavity photon mode and the resultant hybrid states known as polaritons \cite{Carusotto2013,Dovzhenko2018,Basov2021}.}
\label{Fig:interpretation} 
\end{figure*}

The $\sim$7~eV hopping energy of electrons is much higher than the energy of visible photons (0.5-2eV). This \textit{ab initio} theoretical result is compatible with the absence of visible-energy optical transitions (absorption lines) in the transmittance and, more generally, with the transparency of the liquid and supercritical states.

In search for the source of the 2eV energy scale, let us turn our attention to the dielectric constant of SF$_6$ ($\epsilon_r$),  which sets the amplitude of the screening of Coulomb potential. It is linked to the molecular polarizability, $\alpha$, and the fluid density, $N$, through the Clausius-Mossotti relation: 
\begin{equation}
\frac{\epsilon_r-1}{\epsilon_r+2}=\frac{N\alpha}{3\epsilon_0} \quad .
\label{Eq:CM}
\end{equation}

The density of the fluid near the critical point is available thanks to Ref.~\cite{nist} (see Fig.~\ref{Fig:epsilon}a). The extracted inter-molecular distances are shown in Fig.~\ref{Fig:epsilon}b. We have injected the density of fluid  and the polarizability of SF$_6$ ($\alpha$=6.5\AA$^{-3}$ \cite{Hosticka2003,Kita1994}) in  Eq. \ref{Eq:CM} to quantify $\epsilon_r$ of the three phases in our range of interest. The result is shown in Fig.~\ref{Fig:epsilon}c. The experimentally measured values of $\epsilon_r$ \cite{Kita1994} at several temperatures are also shown. The remarkably small $\epsilon_r$ paves the way towards the formation of excitons with a strong binding energy.

A hydrogenic pair of electrons and holes has an effective Bohr radius of: 
\begin{equation}
a^*_{B}=\frac{4 \pi \epsilon_r \epsilon_0 \hbar^2}{e^2}\left(\frac{1}{|m_e|}+\frac{1}{|m_h|}\right) \quad .
\label{Eq:BR}
\end{equation}
With $\epsilon_r < 1.5$, $\frac{1}{m_e} \gg \frac{1}{m_h}$, and $m_e \approx m_0$, Eq.~\ref{Eq:BR} yields an $a^*_{B}\approx$ 1\AA, shorter than the distance between S and F atoms of the same molecule. Therefore, excitons are expected to be tightly bound. The energy of a Frenkel exciton \cite{KNOESTER20031,Zhu2009} is given by:
\begin{equation}
E^F_{\it exc} (d_{eh})=\frac{e^2}{4\pi \epsilon_r \epsilon_0 d_{eh}} \quad .
\label{Eq:Exc1}
\end{equation}
Here, $\epsilon_0$ is the vacuum permittivity, $\epsilon_r$ is the electric permittivity of the medium and $d_{eh}$ is the distance between the electron and the hole. 

We calculated the energy of Frenkel excitons according to Eq. \ref{Eq:Exc1} for the shortest possible $d_{eh}$.  As shown in Fig. \ref{Fig:interpretation}a, the separation between an F atom (locus of an electron) and an S atom (locus of a hole) is $a_1=$1.57 \AA, which is when F and S atoms are on the same molecule. The next possibility is $a_2$. It designates the distance between an S atom and an F atom  located on adjacent molecules. In contrast to $a_1$, $a_2$  varies with the fluid density. Fig. \ref{Fig:interpretation}b shows $E^F_{\it exc}$ for these two types of excitons as a function of temperature, taking the extreme values of   $\epsilon_r$  at each temperature.

The energy separation between $a_1$ and $a_2$ excitons is roughly  2 eV, close to the dip resolved by transmittance experiment.  Thus, exciton formation near the critical point provides a possible explanation. We note that this is not the first case of exciton physics in a liquid. Tightly bound excitons have been detected in liquid Xe above its melting temperature \cite{Steinberger1973}.

The mechanism generating the darkness during a quench is yet to be identified. Fast cooling of the fluid  kept at a constant volume below its critical temperature triggers spinodal decomposition \cite{CAHN1961795,Cahn1965,Elliott1989} of the fluid to its liquid and gaseous components with an intricate nanometric structure. This opens the window to other quantum phenomena. 

Quantum confinement is well-known thanks to research on quantum dots. Clusters of hundreds to thousands of atoms enclosed in a foreign matrix can display optical properties distinct from the bulk material \cite{Alivisatos1996}. Liquid droplets may also become optical micro-cavities capable of morphology-dependent resonances \cite{Giorgini2019}, a cradle for polaritons. The latter are hybrid quasi-particles arising from the interaction of light with electrical-dipole optical transitions \cite{GUILLET2016946,Ribeiro2018,Forndiaz2019}. Following the detection of exciton-photon coupling in semiconductor microcavities three decades ago \cite{Weisbuch1992,Houdre1994,Lidzey1998}, multiple platforms for observing polaritonic phenomena have been reported. Basov \textit{et al.} \cite{Basov2021} have recently listed at least 70 different types of polaritonic light-matter dressing effects. 

Spinodal decomposition is known to generate periodical modulation of the density of the relative composition of the two components of the fluid \cite{CABRAL20181}. This has been established in a variety of systems ranging from copper-nickel alloys \cite{FINDIK2012131} to polymers blends \cite{CABRAL20181}.  The length scale of this modulation depends on physical properties and is generally measured in tens of nanometers. In contrast to nucleation and growth, the decomposition is a gentle, steady process with a steady increase in the local density gradient (see Fig. \ref{Fig:interpretation}c), making it suitable for producing controlled structures at sub-micronic length scales \cite{Germack2013}.

It is tempting to speculate that when this process is triggered by quenching SF$_6$ to its spinodal phase, a liquid-gas morphology emerges which strengthens light-matter interaction.  One possibility is that a gentle gradient of density leads to a multitude of barely separated exciton energy levels. 

\section{\label{sec:CONCLUSION} CONCLUSION}
We found that when near-critical SF$_6$ is quenched, it becomes dark and its transmittance drops by, at least, four orders of magnitude. An optical transition appears in the vicinity of 2~eV. This case of pattern formation in an out-of-equilibrium fluid is distinct from amplified turbidity caused by light scattering at thermal equilibrium in the vicinity of the critical point. The band gap of bulk SF$_6$ is too large, and the vibrational energy is too small to allow the absorption of visible light.  Frenkel excitons with a separation level of $\approx 2$eV are expected to be there. It remains to be understood how the spinodal decomposition caused by the quench triggers darkness across a broad spectral range.  

\section{\label{sec:ack} ACKNOWLEDGMENTS}
We thank Herv\'e Aubin, Daniel Baysens, Beno\^it Fauqu\'e,  Sandrine Ithurria Lhuillier, Arthur Marguerite, Xavier Marie, and Bernard Zappoli for stimulating discussions and critical remarks.  VM acknowledges FAPESP(2018/19420-3). JLJ acknowledges FAPESP(2018/08845-3) and CNPq (310065/2021-6). KB is supported by the Agence Nationale de la Recherche (ANR-19-CE30-0014-04).\\
*\verb|valentina.martelli@usp.br|\\
*\verb|alaska.subedi@polytechnique.edu|\\
*\verb|ricardo.lobo@espci.fr|\\
*\verb|kamran.behnia@espci.fr|\\
\bibliographystyle{unsrt}

%\bibliography{apssamp}
\providecommand{\noopsort}[1]{}\providecommand{\singleletter}[1]{#1}%

%\documentclass[reprint,
%superscriptaddress,
%groupedaddress,
%unsortedaddress,
%runinaddress,
%frontmatterverbose, 
%preprint,
%preprintnumbers,
%nofootinbib,
%nobibnotes,
%bibnotes,
% amsmath,amssymb,
% aps,
%pra,
%prb,
%rmp,
%prstab,
%prstper,
%floatfix, ]{revtex4-2}

%\usepackage{graphicx}% Include figure files
%\usepackage{dcolumn}% Align table columns on decimal point
\newcolumntype{d}[1]{D{.}{.}{#1}}
%\usepackage{bm}% bold math
%\usepackage{ulem}
%\usepackage{hyperref}% add hypertext capabilities
%\usepackage[mathlines]{lineno}% Enable numbering of text and display math
%\linenumbers\relax % Commence numbering lines

%\usepackage[showframe,%Uncomment any one of the following lines to test 
%%scale=0.7, marginratio={1:1, 2:3}, ignoreall,% default settings
%%text={7in,10in},centering,
%%margin=1.5in,
%%total={6.5in,8.75in}, top=1.2in, left=0.9in, includefoot,
%%height=10in,a5paper,hmargin={3cm,0.8in},
%]{geometry}
%\usepackage{xcolor}

%\newcommand{\ddarrows}{\mathbin\downarrow\hspace{-.35em}\downarrow}
%\newcommand{\downuparrows}{\mathbin\downarrow\hspace{-.35em}\uparrow}
%\newcommand{\updownarrows}{\mathbin\uparrow\hspace{-.35em}\downarrow}

%\begin{document}
\title{SUPPLEMENTARY MATERIAL to "Near-critical dark opalescence in out-of-equilibrium SF$_6$"}

\author{Valentina Martelli$^{1,*}$, Amaury Anquetil$^2$, Lin Al Atik$^2$, Julio Larrea Jiménez$^1$, Alaska Subedi,$^{3,*}$, Ricardo P. S. M. Lobo$^{2,*}$ and Kamran Behnia$^{2,*}$}
\affiliation{$^1$ Laboratory for Quantum Matter under Extreme Conditions\\ Institute of Physics, University of São Paulo, São Paulo, Brazil\\
$^2$ Laboratoire de Physique et d'\'Etude des Mat\'eriaux (ESPCI - CNRS - Sorbonne Universit\'e)\\PSL University, Paris, France\\
$^3$ CPHT, CNRS, \'Ecole polytechnique, Institut Polytechnique de Paris, Palaiseau, France
}

\date{\today} 

\maketitle

\section*{Supplementary note 1: Electronic band structure of solid SF$_6$}

%MOVED TO METHODS
%The band structure calculations were performed using the generalized full-potential method as implemented in the {\sc wien2k} package \cite{blaha2020}. Muffin-tin radii of 1.39 and 1.53 a.u.\ for S and F, respectively, were used.  The plane-wave cutoff was set by $RK_{\textrm{max}} = 9$, where $K_{\textrm{max}}$ is the plane-wave cutoff and $R$ is the smallest muffin-tin radius used in the calculations.  $16 \times 16 \times 16$ and $4 \times 4 \times 4$ $k$-point grid was used for the $Im3m$ and $C2/m$ phases, respectively, to perform the Brillouin zone integration in the self-consistent calculations.

\begin{figure}[ht]
\includegraphics[width=1\linewidth]{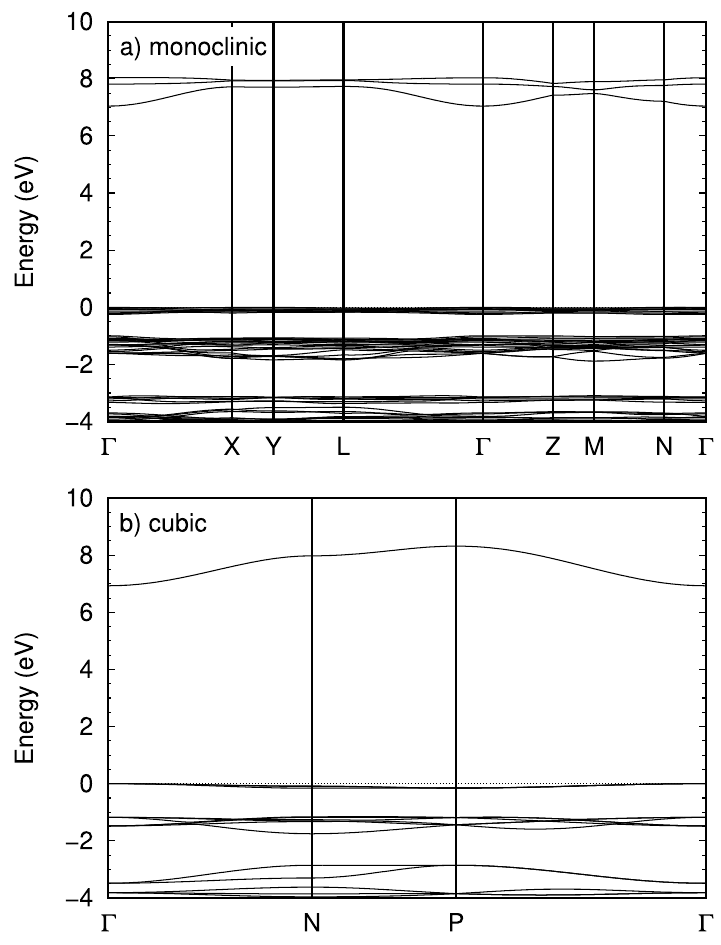} 
']].''']'........................................................]]'\caption{\textbf{Electronic band structure of SF$_6$:} Calculated electronic band structure of solid crystalline SF$_6$ in monoclinic (top) and cubic (bottom) structures. The energy of the valence band edge is set to zero. In both cases, the system is a wide-gap (7 eV) insulator.}
\label{fig:bands} 
\end{figure}

At ambient pressure, SF$_6$ becomes a solid below 209 K. It crystallizes in a body-centred cubic structure with the space group $Im3m$  \cite{Michel1970}. At 95 K, it undergoes a transition to a monoclinic phase with the space group $C2/m$ \cite{Cockcroft1988}. Fig.~\ref{fig:bands} shows our calculations of the band structures of the two phases.% (See the supplement for more detail \cite{SM}). 
In both cases, there is a direct $\sim$7 eV  gap at $\Gamma$  between the valence and conduction bands.  The effective mass for holes ($m_h\approx 4$ $m_0$) is significantly larger than the effective mass for electrons ($m_e\approx m_0$), reflecting the fact that the valence band is flatter than the conduction band. 

\begin{table}[!htbp]
    \caption{Calculated effective masses (in $m_0$) at $\Gamma$ of the highest valence and lowest conduction bands of the cubic $Im3m$ SF$_6$ at different values of the lattice parameter $a$.  The valence bands are triply degenerate at $\Gamma$ and split into doubly degenerate light and   non-degenerate heavy bands along the axial directions, and their masses are denoted by $m_{lh}$ and $m_{hh}$, respectively.  The conduction band is non-degenerate, and its mass is denoted by  $m_e$. The effective masses were calculated by fitting each band around $\pm0.1$ \AA$^{-1}$ of $\Gamma$ with the polynomial $E = E_0 + \hbar^2/2m k^2$. The band gaps of each structure are also given.}
       
    \begin{ruledtabular}
     
      \begin{tabular}{l d{7.2} d{7.2} d{7.2} d{7.2}}
        a (\AA) & \multicolumn{1}{r}{$m_{lh}$} &
        \multicolumn{1}{r}{$m_{hh}$} & \multicolumn{1}{r}{$m_{e}$} &
        \multicolumn{1}{r}{gap (eV)} \\
        \hline
        5.8 &  -6.72   & -35.71  & 0.95 & 6.93   \\ 
        6.2 &  -12.89  & -56.83  & 1.02 & 7.02   \\ 
        6.9 &  -41.09  & -138.23 & 1.16 & 7.18   \\ 
        7.2 &  -67.87  & -205.39 & 1.23 & 7.25   \\
        7.5 &  -111.32 & -309.11 & 1.35 & 7.33   \\
        7.9 &  -220.83 & -528.66 & 1.54 & 7.43
      \end{tabular}
    \end{ruledtabular}
   \label{tab:gap-mass} 
\end{table}

The $Im3m$ cubic crystal has a lattice parameter of 5.795 \AA. This corresponds to an intermolecular distance of 5.02 \AA. We have computed the evolution of the electronic band structure of SF$_6$ with increasing lattice parameter. The results are listed in Table \ref{tab:gap-mass}. The band gap and the effective masses steadily increase with the enhancement of the distance between molecules. The change in electron mass with increasing lattice parameter is modest. On the other hand, the hole mass increases much more drastically. The difference can be tracked to the different origins of the electrons and hole states. The conduction band mainly originates from a sulfur $3s$ orbital, while the valence bands are predominantly associated with fluorine $2p$ orbitals. The large magnitude of the band gap indicates that electron sharing between neighbouring molecules is very weak.

The $Im3m$ cubic and $C2/m$ monoclinic phases have one and three formula units per primitive cell, respectively.  This results in one and three lowest-lying conduction bands for the $Im3m$ and $C2/m$ phases, respectively. In both phases, the lowest-lying conduction bands have mainly sulfur $3s$ character, although there is some mixing with the fluorine $2p$ orbitals due to covalency. The highest-lying valence bands have predominantly fluorine $2p$ character in both phases.  However, there is noticeably more $3s$ character in the valence bands of the $C2/m$ phase, presumably because of mixing due to lower symmetry. The structural parameters of the two phases were obtained from powder neutron diffraction \cite{Cockcroft1988}. 

\section*{Supplementary methods: Thermal gradient inside the chamber}
%The IR camera is set on $T_{min}$ at a value ranging from 25$^{\circ}$C to 54$^{\circ}$C and $T_{max}$ is set to a fixed value of 56$^{\circ}$C. The image is displayed in colour codes ranging from red to blue. The screen is entirely red when the fluid is at a temperature higher than $T_{max}$. When the cooling process begins, the image gradually fades away when the colder side of the chamber reaches $T\leq T_{min}$ (Fig. \ref{fig:supplIR1}). Keeping constant $T_{max}$ and changing $T_{min}$ for different cooling processes, we can relate the cold (and hot) side of the gradient with respect to the chamber temperature.

%\begin{figure}[h]
%\includegraphics[width=0.9\linewidth]{IR_intro.jpg} 
%\caption{\label{fig:supplIR1} Example of an IR-image for the %fluid in the critical state ($T>T_C$).}
%\end{figure}

\begin{figure}[h]
\includegraphics[width=1\linewidth]{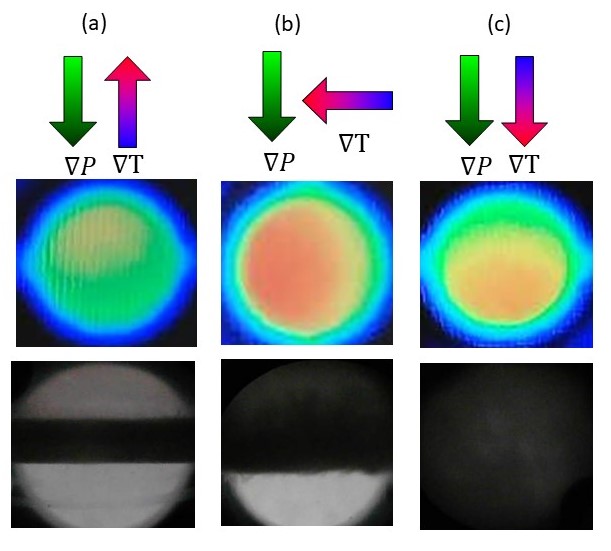} 
\caption{\label{fig:supplIRcomparison} Optical and IR frames for the three different configurations for a cooling down process with rate $\sim$ 0.3K/min.}
\end{figure}

Comparing cooling down processes observed in similar conditions with the optical and IR cameras, we verified that within our resolution, there is no correlation between the region where the black pattern appears and the temperature distribution. 
The IR pictures are in colour codes where green points to cold and red/yellow towards the hot end. The gradient follows the chamber's orientation and agrees with the position of the cooling pipes (Fig. \ref{fig:supplIRcomparison}. The temperature sensor measuring $T_{Pt}$, installed into the lateral cavity, is placed on the cold end of the chamber during the cooling out-of-equilibrium process.

%In figure \ref{fig:supplfigIRT1}, we can estimate the evolution of the cold side of the chamber as a function of the minimum temperature set in the IR-camera, just observing when the IR frame begins to fade away. For both the high cooling rate (0.3K/s) and low cooling rate (0.03K/s), there is a linear relation, pointing out that the chamber's temperature directly relates to the cold end of the gradient. On the other end, when the same relation is investigated at the moment the IR frame becomes fully dark (see Fig. \ref{fig:supplfigIRT2}), an estimation of the hot end of the gradient can be done. We observe that a difference higher than 10K can exist between the hot end of the gradient and the read temperature by the platinum sensor. That difference becomes smaller (but still sizable) for a low cooling rate (open symbols). 

%\begin{figure}[h]
%\includegraphics[width=1\linewidth]{IR-figT1.jpg} 
%\caption{\label{fig:supplfigIRT1} Cold side of the IR frame as a function of the read temperature on chamber temperature $T_Pt$ when the IR image begins to fade away.}
%\end{figure}

%\begin{figure}[h]
%\includegraphics[width=1\linewidth]{IR-figT2.jpg} 
%\caption{\label{fig:supplfigIRT2}  Hot side of the IR frame as a function of the read temperature on chamber temperature $T_Pt$ when the IR image fully disappears.}
%\end{figure}
\section*{Supplementary note 2: Comparison of the temperature evolution in the three configurations}
\begin{figure}[ht]
\includegraphics[width=1\linewidth]{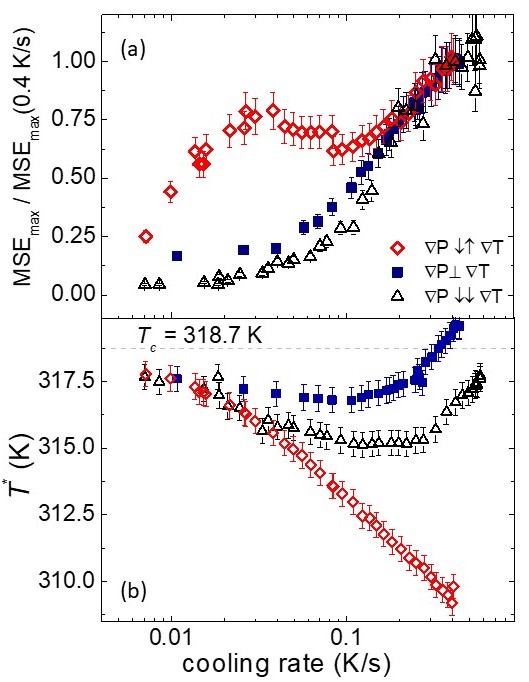} 
\caption{\label{fig:MSE-T-max}\textbf{Comparison of the three configurations:} (a) The maximum in the MSE index is plotted against the cooling rate for the three configurations. (b) The chamber temperature $T^*$ at which the MSE reaches its maximum as a function of the cooling rate. Error bars are estimated to be 10}
% based on the average results obtained repeating the quenching under same conditions. }
\end{figure}

Despite the identical cooling protocol, salient features of the collected data differ in the three cases. Figure \ref{fig:MSE-T-max}(a) shows the normalized intensity of the peak as a function of the cooling rate, for the three configurations. Blackness is reduced with a decreasing cooling rate.  The evolution is similar for perpendicular ($\nabla P$ $\perp$ $\nabla T$) and parallel ($\nabla P$ $\ddarrows$ $\nabla T$) configurations. A different trend is visible in the anti-parallel case ($\nabla P$ $\downuparrows$ $\nabla T$).
%compatible with that observed in Fig.~\ref{fig:antiparallel}. 
As the cooling rate is reduced, the black band, despite becoming more transparent, occupies a larger area, leading to a plateau at around 0.02~K/s, before approaching zero at lower cooling rates.

Fig.~\ref{fig:MSE-T-max}(b) shows the evolution of $T^*$, the temperature at which the MSE peaks as a function of the cooling rate. Note that this is the temperature of a sensor outside the chamber and not the non-uniform temperature of the fluid. Nevertheless, given the identity of the cooling protocol, the variation of $T^*$ reveals a genuine difference in the cooling dynamics, presumably due to the difference in the relative weights of various thermalization mechanisms in the three configurations. When the cooling rate is slower than 0.01 K/s, $T^*$ becomes identical for the three configurations and saturates to slightly below the critical temperature, suggesting that turbidity peaks indeed at the critical temperature, when the thermodynamic equilibrium is kept. The 1 K difference between the measured temperature and $T_c$ can be attributed to experimental inaccuracy as a result of the absence of any sensor inside the chamber. As the cooling rate increases, $T^*$ becomes significantly lower than the critical temperature.
 %suggesting that the maximum turbidity in this pure fluid pushed out of equilibrium may happen close to the Frenkel line. 
 The anti-parallel case  ($\nabla$P $\downuparrows$ to $\nabla$T) is the most striking. For high cooling rates, $T^*$ becomes $\sim$10 K lower than $T_c$, pointing to non-trivial dynamics well below the boiling and the Widom lines.  

\section*{Supplementary note 3: Selected videos}
A selection of videos obtained during the experiments is provided with the supplementary material. Table \ref{tab:videos} reports some details to identify the configuration adopted for each video. 

\begin{table*}[]
\begin{center}
\begin{tabular}{ |c|c|l| } 
 \hline
CONFIGURATION & COOLING RATE & FILE NAME \\ 
 \hline
     $\nabla P$ $\downuparrows$ $\nabla T$  & 0.4~K/s & C1\_antiparallel\_FASTcooling.mp4\\ 
  $\nabla P$ $\downuparrows$ $\nabla T$ & 0.03~K/s & C1\_antiparallel\_SLOWcooling.mp4 \\ 
                                       \hline
$\nabla T$ $\perp$ $\nabla P$ & 0.4~K/s & C2\_perpendicular\_FASTcooling.mp4 \\ 
 $\nabla T$ $\perp$ $\nabla P$          & 0.025~K/s& C2\_perpendicular\_SLOWcooling.mp4 \\ 
                                        \hline
  $\nabla T$ $\ddarrows$ $\nabla P$      &0.4~K/s& C3\_parallel\_FASTcooling.mp4 \\ 
  $\nabla T$ $\ddarrows$ $\nabla P$     & 0.04~K/s & C3\_parallel\_SLOWcooling.mp4  \\ 
%  $\nabla T$ $\ddarrows$ $\nabla P$   & warming up  & v10\_parallel\_warmingUP.avi \\ 
 
 \hline
\end{tabular}
\caption{\label{tab:videos} Details of the videos of the Supplementary material for the three configurations.}
\end{center}
\end{table*}

\section*{Supplementary note 4: Turbidity near the critical point} 
According to Puglielli and Ford \cite{Puglielli1970}, incorporating the Rayleigh and the Brillouin contributions to the light scattering will lead to the following expression for the intensity of light scattering per unit length:
\begin{equation}
I (k)= \frac{\pi^2}{\lambda^4}\rho^2\left[\frac{\partial n}{\partial \rho}\right]^2k_BT \beta \frac{sin^2\Phi}{1+ (q\xi)^2}
\label{Eq:I}
\end{equation}
Here, $k$ is the wavevector of the incident wave, $\lambda=\frac{2\pi}{k}$ is it wavelength, $q$ is the change in the wavevector of the incident and scattered wave, $\Phi$ is the  angle,  $\rho$ is the density, $n$ is the refractive index and $\beta$ is the isothermal compressibility. Using the Lorenz-Lorentz relation \cite{Kragh_2018} between the refractive index and the density and integrating over all angles, one obtains an expression for OZ turbidity \cite{Puglielli1970,Lecoutre2009} proportional to the diverging compressibility ($\beta$):
\begin{equation}
\tau_{OZ}= F(a)\frac{\pi^3}{\lambda^4}\left[\frac{(n^2-1)(n^2+2)}{3}\right]^2k_BT_c \beta 
\label{Eq:turbidity}
\end{equation}
Here, $a= 2(k\xi)^2$ and:
\begin{equation}
F(a)=\left(\frac{2a^2+2a+1}{a^3}\right)ln(1+2a)-2\left(\frac{1+a}{a^2}\right).
\label{Eq:F}
\end{equation}

The validity of Supplementary equation (\ref{Eq:turbidity}) near the critical point of SF$_6$ was experimentally confirmed on Earth \cite{Puglielli1970} (for $(T-T_c)>0.1$ K) and by very precise micro-gravity experiments much closer to the critical temperature \cite{Lecoutre2009}. 
%In this picture, scattering  diverges when the critical point is approached from the supercritical side driven by the divergent thermal compressibility. It predicts a turbidity of about 0.02 cm$^{-1}$ when the temperature is 1 K off the critical temperature. This is a far cry from the blackness detected in out-of-equilibrium conditions.  Moreover, a key experimental observation, that proximity to the critical point along a adiabatic trajectory weakens the darkness instead of amplifying the turbidity does not fit into this picture which is  based on the way visible phonons are scattered either by elastic Rayleigh  scattering (which brings $\frac{1}{\lambda^4}$) or by inelastic Brillouin scattering (which is the origin of $\frac{1}{(q\xi)^2}$). 

\section*{Supplementary discussion: Previous numerical simulations} Zappoli and collaborators simulated the behavior of a supercritical fluid with Earth gravity enclosed in a finite volume, either heated from a side ($\nabla T \perp \nabla P$) \cite{zappoli1996thermoacoustic_buoyancy} or from the bottom ($\nabla T \parallel \nabla P$) \cite{amiroudine2004thermoconvective_instabilities}. These two cases are analogous to two of our configurations [Fig. 2 (c) and (d)].  

In the first case ($\nabla T \perp \nabla P$), they found that the combination of the buoyancy and the piston effects generates a plume in density and in temperature in an upper corner of the volume. In the second case ($\nabla T \parallel \nabla P$), they found that there is  a concomitant variation of density and temperature both at the top and bottom of the fluid. 

 It is tempting to make a qualitative link between these simulations and our observations. In the first case, we see a black pattern appearing from the top part of the chamber, whereas in the second case, blackness covers the entire volume (both from the top and the bottom). The temperature difference in these  simulations  \cite{zappoli1996thermoacoustic_buoyancy, amiroudine2004thermoconvective_instabilities} is two orders of magnitude smaller than in our experiment. Therefore, this qualitative similarity should be taken with a grain of salt. However, it strengthens the plausibility of a correlation between blackness and enhanced density.

%\bibliography{apssamp}

\providecommand{\noopsort}[1]{}\providecommand{\singleletter}[1]{#1}%

\section*{\label{sec:cor} Corresponding authors
}
Correspondence should be addressed to Kamran Behnia, Valentina Martelli, Alaska Subedi, and Ricardo Lobo.

*\verb|valentina.martelli@usp.br|\\
*\verb|alaska.subedi@polytechnique.edu|\\
*\verb|ricardo.lobo@espci.fr|\\
*\verb|kamran.behnia@espci.fr|\\
%\end{document}

\end{document}